\documentclass[12pt]{article}
\usepackage{amsmath}
\usepackage{graphicx}
\usepackage{epstopdf}
\usepackage[russian]{babel}
\usepackage{amsfonts}
\usepackage{amssymb}
\usepackage{stmaryrd}
\usepackage{amscd}
\newcommand{\bc}{\begin{center}}
\newcommand{\ec}{\end{center}}
\newcommand{\be}{\begin{equation}}
\newcommand{\ee}{\end{equation}}
\newcommand{\bea}{\begin{eqnarray}}
\newcommand{\eea}{\end{eqnarray}}
\newcommand{\ba}{\begin{array}}
\newcommand{\ea}{\end{array}}
\newcommand{\edc}{\end{document}}

\def\l{\lambda}

\begin{document}
\thispagestyle{empty}

УДК 517.98

\begin{center}
{\bf \large {Крайность трансляционно-инвариантной меры Гиббса для НС-модели на дереве Кэли}}
\end{center}

\begin{center}
У.А.Розиков\footnote{Институт математики, ул. Дурмон йули, 29, Ташкент, 100125, Узбекистан.\\
E-mail: rozikovu@yandex.ru}, Р.М.Хакимов\footnote{Институт математики, ул. Дурмон йули, 29, Ташкент, 100125, Узбекистан.\\
E-mail: rustam-7102@rambler.ru}
\end{center}

В данной статье изучается крайность трансляционно-инвариантной меры Гиббса для Hard-Core (HC) модели на дереве Кэли. Известно, что трансляционно-инвариантная мера для этой модели единственна. Дано новое доказательство этого утверждения и найдены области крайности этой меры на дереве Кэли. \\

\textbf{Ключевые слова}: дерево Кэли, допустимая конфигурация, НС-модель, мера Гиббса, трансляционно-инвариантные меры, крайность меры.\

\section{Введение}\

 Для достаточно широкого класса гамильтонианов известно, что множество всех предельных мер Гиббса (соответствующих данному гамильтониану) образует непустое выпуклое компактное подмножество в множестве всех вероятностных мер (см. например \cite {Si}), и каждая точка этого выпуклого множества однозначно разлагается по его крайним точкам. В связи с этим особый интерес представляет описание всех крайних точек этого выпуклого множества, т. е. крайних мер Гиббса.

В работе \cite {7} была доказана единственность трансляционно-инвариантной меры Гиббса (ТИМГ) и не
единственность периодических мер Гиббса для НС-модели. Также в \cite{7} (соответственно в \cite{Mar}) найдено достаточное условие на параметры НС-модели, при котором ТИМГ является не крайней (соответственно крайней). В работе \cite{RKh} изучены слабо периодические меры Гиббса для
HC-модели для нормального делителя индекса два и при некоторых условиях на
параметры показана единственность слабо периодической меры Гиббса, а в работе \cite{XR} доказана единственность (трансляционно-инвариантность) слабо периодической меры Гиббса для HC-модели при любых значениях параметров. В \cite{KhR} доказано существование слабо периодических (не периодических) мер Гиббса для HC-модели для нормального делителя индекса четыре на некоторых инвариантах при некоторых условиях на параметры.

Для ознакомления с другими свойствами НС-модели (и их обобщения) на дереве Кэли см. Главу 7 монографии \cite{Rb}.

В данной работе изучается крайность ТИМГ для HC-модели на дереве Кэли порядка $k\geq2$.  Мы применяем методы из работы \cite{MSW}. Дано новое доказательство единственности ТИМГ при всех $k\geq 2$.  Найдено условие крайности ТИМГ. Кроме того, при некоторых $k$ даны явные оценки на параметр модели. Наш результат улучшает результаты работ \cite{7}, \cite{Mar} при $k\leq 18$.

\section{Определения и известные факты}\

Дерево Кэли $\Gamma^k$ порядка $ k\geq 1- $ бесконечное дерево, т.е. граф без циклов, из
каждой вершины которого выходит ровно $k+1$ ребро. Пусть
$\Gamma^k=(V,L,i)$, где $V-$ есть множество вершин $\Gamma^k$, $L-$
его множество ребер и $i-$ функция инцидентности,
сопоставляющая каждому ребру $l\in L$ его концевые точки $x, y \in
V$. Если $i (l) = \{ x, y \} $, то $x$ и $y$ называются  {\it
ближайшими соседями вершины} и обозначается $l = \langle
x,y\rangle $. Расстояние $d(x,y), x, y \in V$ на дереве Кэли
определяется формулой
$$
d (x, y) = \min \{d | \exists x=x_0,x_1, \dots, x_{d-1},
x_d=y\in V \ \ \mbox {такие, что} \ \ \langle x_0,x_1\rangle,\dots, \langle x_
{d-1}, x_d\rangle\}.$$

Для фиксированного $x^0\in V$ обозначим
$$ W_n =\{x\in V | \ d (x, x^0) =n \}, \  V_n = \{x\in V | \ d (x, x^0) \leq n\}.$$
Для $x\in W_{n}$ обозначим (множество прямых потомков вершины $x$)
$$ S(x)=\{y\in{W_{n+1}}:d(x,y)=1\}.$$

Известно, что существует взаимнооднозначное соответствие между
множеством $V$ вершин дерева Кэли порядка $k\geq 1$ и группой
$G_k$, являющейся свободным произведением $k+1$ циклических групп
второго порядка с образующими $a_1,...,a_{k+1}$, соответственно
(см. Главу 1 \cite{Rb}). Поэтому можно отождествлять множество $V$ c множеством $G_k$.

Пусть $\Phi=\{0,1\}$ и $\sigma\in\Phi^V$-конфигурация, то есть
$\sigma=\{\sigma(x)\in \Phi: x\in V\}$, где $\sigma (x)=1$
означает, что вершина $x$ на дереве Кэли занятая, а $\sigma (x)=0$
означает, что она свободная. Конфигурация $\sigma$ называется
допустимой, если $\sigma (x)\sigma (y)=0$ для любых соседних
$\langle x,y \rangle $ из $V  (V_n $ или $W_n$, соответственно ) и обозначим множество таких
конфигураций через $\Omega$ ($\Omega_{V_n}$ и $\Omega_{W_n}).$
Ясно, что $\Omega\subset\Phi^V.$

Для $\sigma_n\in\Omega_{V_n}$ положим
$$\#\sigma_n=\sum\limits_{x\in V_n}{\mathbf 1}(\sigma_n(x)\geq 1)$$
число занятых вершин в $\sigma_n$.

Пусть $z:x\mapsto z_x=(z_{0,x}, z_{1,x}) \in R^2_+$ векторнозначная функция на $V$. Для $n=1,2,\ldots$ и $\l>0$
рассмотрим вероятностную меру $\mu^{(n)}$ на $\Omega_{V_n}$,
определяемую как
\begin{equation}\label{rus2.1}
\mu^{(n)}(\sigma_n)=\frac{1}{Z_n}\lambda^{\#\sigma_n} \prod_{x\in
W_n}z_{\sigma(x),x}.
\end{equation}
Здесь $Z_n$-нормирующий делитель:
$$
Z_n=\sum_{{\widetilde\sigma}_n\in\Omega_{V_n}}
\lambda^{\#{\widetilde\sigma}_n}\prod_{x\in W_n}
z_{{\widetilde\sigma}(x),x}.
$$
Говорят, что вероятностная мера $\mu^{(n)}$ является
согласованной, если $\forall$ $n\geq 1$ и
$\sigma_{n-1}\in\Omega_{V_{n-1}}$:
\begin{equation}\label{rus2.2}
\sum_{\omega_n\in\Omega_{W_n}}
\mu^{(n)}(\sigma_{n-1}\vee\omega_n){\mathbf 1}(
\sigma_{n-1}\vee\omega_n\in\Omega_{V_n})=
\mu^{(n-1)}(\sigma_{n-1}).
\end{equation}
В этом случае существует единственная мера $\mu$ на $(\Omega,
\textbf{B})$ такая, что для всех $n$ и $\sigma_n\in
\Omega_{V_n}$
$$\mu(\{\sigma|_{V_n}=\sigma_n\})=\mu^{(n)}(\sigma_n),$$
где $\textbf{B}-$$\sigma$-алгебра, порожденная цилиндрическими
подмножествами $\Omega$.

\textbf{Определение 1.} Мера $\mu$, определенная формулой
(\ref{rus2.1}) с условием согласованности (\ref{rus2.2}),
называется HC-\textit{мерой Гиббса} с $\lambda>0$,
\textit{соответствующей функции} $z:\,x\in V
\setminus\{x^0\}\mapsto z_x$.

В следующем утверждении сформулировано условие на $z_x$, гарантирующее
согласованность меры $\mu^{(n)}$.

\textbf{Утверждение 1.}\cite{7}. Вероятностные меры
$\mu^{(n)}$, $n=1,2,\ldots$, заданные формулой (\ref{rus2.1}),
согласованны тогда и только тогда, когда для любого $x\in V$ имеет
место следующее равенство
\begin{equation}\label{rus2.3}
z_x^{'}=\prod_{y \in S(x)}(1+\lambda z_y^{'})^{-1},
\end{equation}
где $z^{'}_x=z_{1,x}/z_{0,x}$,  $\lambda>0-$параметр.

Пусть $\widehat{G}_k-$подгруппа группы $G_k$.

\textbf{Определение 2}. Совокупность величин $z=\{z_x,x\in G_k\}$
называется $ \widehat{G}_k$-периодической, если  $z_{yx}=z_x$ для
$\forall x\in G_k, y\in\widehat{G}_k.$

$G_k-$периодическая совокупность называется трансляционно-инвариантной.

\textbf{Определение 3}. Мера $\mu$ называется
$\widehat{G}_k$-периодической, если она соответствует
$\widehat{G}_k$-периодической совокупности величин $z$.

\textbf{Теорема 1.} \cite{RKh}. Для любого нормального делителя
${\mathcal G}\subset G_k$ всякая $\mathcal G$-периодическая мера
Гиббса НС-модели является либо трансляционно-инвариантной,
либо  $G^{(2)}_k$-периодической мерой Гиббса, где
$G^{(2)}_k$-подгруппа, состоящая из слов четной длины.

\textbf{Утверждение 2.} $\forall \lambda>0$ и $k\geq2$ ТИМГ для
HC-модели единственна.

\textbf{Доказательство.} Заметим, что ТИМГ соответствует решению (\ref{rus2.3}) вида $z_x'=z$ для всех $x\in V$, где
$z$ удовлетворяет уравнению
\begin{equation}\label{rus2.4}
z={1\over (1+\lambda z)^k}.
\end{equation}
Покажем, что (\ref{rus2.4}) имеет единственное положительное решение при любых значениях $\lambda>0$ и $k\geq2$.
Действительно, перепишем уравнение (\ref{rus2.4}) в виде
$$f(z)=\lambda^kz^{k+1}+k\lambda^{k-1}z^k+...+z-1=0.$$
Ясно, что $f(0)=-1$ и $f(+\infty)=+\infty$. Поэтому уравнение $f(z)=0$ имеет не менее одного решения. Кроме того, по теореме о количестве положителных корней многочлена (см. стр.28, \cite{Pra}) уравнение $f(z)=0$ имеет не более одного положительного решения. Следовательно, уравнение (\ref{rus2.4}) имеет единственное положительное решение. Утверждение доказано.

\textbf{Замечание 1.} В работе \cite{7} единственность решения уравнения (\ref{rus2.4}) была доказана, используя монотонность функции в правой части этого уравнения.

Пусть $\mu^{*}$ обозначает ТИМГ, соответствующую решении уравнения (\ref{rus2.4}).

Известно (\cite{Rb}), что при
\begin{equation}\label{rus2.18}
\lambda\leq\lambda_{cr}=\lambda_{cr}(k)={k^k\over(k-1)^{k+1}}
\end{equation}
$G^{(2)}_k$-периодическая HC-мера совпадает с $\mu^{*}$, а при $\lambda>\lambda_{cr}$ существуют не менее трех
$G^{(2)}_k$-периодических мер Гиббса, одна из которых является $\mu^{*}$.\

\section{Крайность трансляционно-инвариантной меры Гиббса}\

Напомним (см. \cite{Si}), что мера Гиббса $\mu$ (как элемент выпуклого множества) называется крайней,
если $\mu\ne s\nu+(1-s)\nu'$ для различных мер Гиббса $\nu, \nu'$ и $0<s<1$.

Из работ \cite{7} и \cite{Mar} известны следующие теоремы и предложение:

\textbf{Теорема 2.} \cite{7}. При $k\geq2$ и
\begin{equation}\label{rus2.60}
\lambda>{1\over \sqrt{k}-1}\left({\sqrt{k}\over \sqrt{k}-1}\right)^k
\end{equation}
ТИМГ $\mu^{*}$ является не крайней.\

\textbf{Предложение 1.} \cite{7}. Если для данных $k$ и $\lambda_0$ мера $\mu^{*}$ является не крайней,
то она является не крайней при всех $\lambda>\lambda_0$.

\textbf{Теорема 3.} \cite{Mar}. При $k\geq2$ и $\lambda=1$
ТИМГ $\mu^{*}$ является крайней.\

Из Предложения 1 и Теоремы 3 получаем

\textbf{Следствие 1.} При $k\geq2$ и $\lambda\in (0,1]$ мера
$\mu^{*}$ является крайней.

Основным результатом данной работы является следующая

\textbf{Теорема 4.} При $k\geq2$ и $\lambda\in (0,\lambda_*)$
ТИМГ $\mu^{*}$ является крайней, где
\begin{equation}\label{l*}
\lambda_*=\lambda_*(k)={1\over t_*^k}\left({1\over t_*}-1\right),
\end{equation}
и $t_*\in (0,1)$ является единственным решением уравнения
\begin{equation}\label{z0}
t^{k+1}-k t^2+(2k-1)t-k+1=0.
\end{equation}

\textbf{Доказательство.}
Мы используем методы из работы \cite{MSW} (см. также Лемму 5.7 из \cite{KR}).
Рассмотрим цепь Маркова с состояниями $\{0,1\}$ и матрицу
$\mathbf{P_{\mu^*}}$ вероятностных переходов $P_{ij}$, определенную данной
ТИМГ $\mu^*$ следующим образом:
\begin{equation}\label{rus2.7}
\mathbf{P_{\mu^*}}=\left(%
\begin{array}{cccccc}
 {1\over 1+\lambda z} &  {\lambda z\over1+\lambda z} \\[2 mm]
 1 &  0  \\
  \end{array}%
\right),
\end{equation}
где $z$ есть единственное решение уравнения (\ref{rus2.4}). Заметим, что $z\in (0,1)$.

Приведем необходимые определения из работы \cite{MSW}.
Если удалить произвольное ребро $\langle x^0, x^1\rangle=l\in L$ из дерева Кэли $\Gamma^k$, то оно разбивается на две компоненты $\Gamma^k_{x^0}$ и $\Gamma^k_{x^1}$, каждая из которых называется полубесконечным деревом или полудеревом Кэли.

 Рассмотрим конечное полное поддерево $\mathcal T$, которое содержит все начальные точки полудерева $\Gamma^k_{x^0}$. Граница $\partial \mathcal T$ поддерева $\mathcal T$ состоит из ближайших соседей его вершин, которые лежат в $\Gamma^k_{x^0}\setminus \mathcal T$. Мы отождествляем поддерево $\mathcal T$ с множеством его вершин. Через $E(A)$ обозначим множество всех ребер $A$ и $\partial A$.

В \cite{MSW} ключевыми являются две величины  $\kappa$ и $\gamma$. Оба являются свойствами множества мер Гиббса $\{\mu^\tau_{{\mathcal T}}\}$, где граничное условие $\tau$ фиксировано и $\mathcal T$ является произвольным, начальным, полным, конечным поддеревом $\Gamma^k_{x^0}$. Для данного начального поддерева $\mathcal T$ дерева $\Gamma^k_{x^0}$ и вершины $x\in\mathcal T$ мы будем писать $\mathcal T_x$ для (максимального) поддерева $\mathcal T$  с начальной точкой в $x$. Когда $x$ не является начальной точкой $\mathcal T$, через $\mu_{\mathcal T_x}^s$ обозначим меру Гиббса, в которой "предок"  $x$ имеет спин $s$ и конфигурация на нижней границе ${\mathcal T}_x$ (т.е. на $\partial {\mathcal T}_x\setminus \{\mbox{предок}\ \ x\}$) задается через $\tau$.

Для двух мер $\mu_1$ и $\mu_2$ на $\Omega$ через $\|\mu_1-\mu_2\|_x$ обозначим расстояние по норме
$$\|\mu_1-\mu_2\|_x={1\over 2}\sum_{i=0}^1|\mu_1(\sigma(x)=i)-\mu_2(\sigma(x)=i)|.$$
Пусть $\eta^{x,s}$ есть конфигурация $\eta$ со спином в $x$, равным $s$.

Следуя \cite{MSW}, определим
$$\kappa\equiv \kappa(\mu)={1\over2}\max_{i,j}\sum_{l=0}^1|P_{il}-P_{jl}|;$$
$$\gamma\equiv\gamma(\mu)=\sup_{A\subset \Gamma^k}\max\|\mu^{\eta^{y,s}}_A-\mu^{\eta^{y,s'}}_A\|_x,$$
где максимум берется по всем граничным условиям $\eta$, всеми $y\in \partial A$, всеми соседями $x\in A$ вершины $y$ и всеми спинами $s, s'\in \{0,1\}$.

Достаточным условием крайности  меры Гиббса $\mu$ является $k\kappa(\mu)\gamma(\mu)<1$.

 Используя (\ref{rus2.7}), при $i\neq j$ получим $\kappa={\lambda z\over 1+\lambda z}$.
Из работы \cite{MSW}(стр.151, Теорема 5.1.) известно, что для HC-модели справедлива оценка: $\gamma\leq{\lambda\over \lambda+1}$. Значит, для крайности меры $\mu^{*}$ достаточно выполнение неравенства
\begin{equation}\label{rus2.8}
k\cdot{\lambda z\over 1+\lambda z}\cdot{\lambda\over \lambda+1}<1.
\end{equation}
Из уравнения (\ref{rus2.4}) находим
\begin{equation}\label{lz}
\lambda=\lambda(z)={1\over z}\left({1\over \sqrt[k]{z}}-1\right).
\end{equation}
Подставляя это выражение в (\ref{rus2.8}), получим
\begin{equation}\label{zk}
z\sqrt[k]{z}-k \sqrt[k]{z^2}+(2k-1)\sqrt[k]{z}-k+1>0.
\end{equation}
Обозначим $t=\sqrt[k]{z}$. Ясно, что $z\in (0,1)$. Отсюда имеем $t\in (0,1)$.
Из (\ref{zk}) получим
\begin{equation}\label{t0}
\varphi(t)=t^{k+1}-k t^2+(2k-1)t-k+1>0.
\end{equation}
Докажем, что неравенство (\ref{t0}) верно при всех
$t\in (t_*,1)$, (где $t_*-$решение уравнения (\ref{z0}), т.е. $\varphi(t)=0$).
Имеем
$$
\varphi'(t)=(k+1)t^{k}-2k t+(2k-1)>\min_{t\in (0,1)}\varphi'(t)=\varphi'\left(\sqrt[k-1]{{2\over k+1}}\right)=$$
$$=2k-1-2(k-1)\sqrt[k-1]{{2\over k+1}}>2k-1-2(k-1)=1>0.
$$
Следовательно, функция $\varphi(t)$ при $t\in (0,1)$ возрастает и $\varphi(0)=1-k<0$, $\varphi(1)=1>0$.
Поэтому $\varphi(t)=0$ имеет единственное решение $t_*\in (0,1)$. Следовательно, неравенство $\varphi(t)>0$
имеет решение только вида $(t_*,1)$.

Заметим, что функция $\lambda(z)=\lambda(t^k)$ (см. (\ref{lz})) монотонно убывающая. Значит, из $t\in(t_*,1)$
следует $\lambda\in (0,\lambda_*)$. Теорема доказана.

\textbf{Замечание 2.} Так как при $k \geq 4$ уравнение (\ref{z0}) не
решается в радикалах, то не легко получить явное значение $\lambda_*(k)$.
Но при каждом фиксированном значении $k\geq 2$ компьютерный анализ дает
приблизительное значение $\lambda_*(k)$.
Приведем эти значения при некоторых $k$:

$$\lambda_*(2)={1\over3}\sqrt[3]{388+12\sqrt{69}}+{52\over 3\sqrt[3]{388+12\sqrt{69}}}+{7\over3}\approx 7.159191247,$$
$$\lambda_*(3) \approx 3.771210223, \ \ \lambda_*(4)\approx 2.745407267, \ \ \lambda_*(5)\approx 2.242274271,$$
$$\lambda_*(18) \approx 1.015307312, \ \ \lambda_*(19)\approx 0.9900721365,$$
$$\lambda_*(2016)\approx 0.2680659445, \ \ \lambda_*(3000)\approx  0.2497937373.$$
Следовательно, Теорема 4 улучшает результат Следствии 1 при всех $k\leq 18$.

\textbf{Замечание 3.} Делим обе стороны уравнения (\ref{z0}) на $k$ и рассмотрим $k\to \infty$.
Тогда при $t\in (0,1)$ уравнение "стремится" \, к $t^2-2t+1=0$, т.е. $t_*\to 1$ при $k\to \infty$.
Следовательно, $\lim_{k\to \infty}\lambda_*(k)=0$.

\textbf{Замечание 4.} Пусть $\widetilde{\lambda}_k=\lambda_*(k)-\lambda_{cr}(k), \ k\geq 2$, где $\lambda_{cr}$ определено в неравенстве (\ref{rus2.18}). Тогда $\widetilde{\lambda}_2=3.159191247, \ \widetilde{\lambda}_3=2.083710221, \ \widetilde{\lambda}_4\approx1.691909325, \ \widetilde{\lambda}_5\approx1.479334818$.

\end{document}